\documentclass[aps,prc,preprint,showpacs,showkeys,floatfix,nofootinbib,superscriptaddress,a4paper]{revtex4}

\newcommand{\rd}{\mbox{{\rm d}}}

\usepackage{graphicx,amssymb,amsmath}
\usepackage{dcolumn}
\usepackage{color}

\begin{document}

\title{Study of the $pd\to\,^3\textrm{He}\,K^+K^-$
and $pd\to\,^3\textrm{He}\,\phi$ reactions close to threshold}

\author{F.~Bellemann}
\affiliation{Helmholtz-Institut f\"{u}r Strahlen- und Kernphysik der
Universit\"{a}t Bonn, 53115 Bonn, Germany}

\author{A.~Berg}
\affiliation{Helmholtz-Institut f\"{u}r Strahlen- und Kernphysik der
Universit\"{a}t Bonn, 53115 Bonn, Germany}

\author{J.~Bisplinghoff}
\affiliation{Helmholtz-Institut f\"{u}r Strahlen- und Kernphysik der
Universit\"{a}t Bonn, 53115 Bonn, Germany}

\author{G.~Bohlscheid}
\affiliation{Helmholtz-Institut f\"{u}r Strahlen- und Kernphysik der
Universit\"{a}t Bonn, 53115 Bonn, Germany}

\author{J.~Ernst}
\affiliation{Helmholtz-Institut f\"{u}r Strahlen- und Kernphysik der
Universit\"{a}t Bonn, 53115 Bonn, Germany}

\author{C.~Henrich}
\affiliation{Helmholtz-Institut f\"{u}r Strahlen- und Kernphysik der
Universit\"{a}t Bonn, 53115 Bonn, Germany}

\author{F.~Hinterberger}
\affiliation{Helmholtz-Institut f\"{u}r Strahlen- und Kernphysik der
Universit\"{a}t Bonn, 53115 Bonn, Germany}

\author{R.~Ibald}
\affiliation{Helmholtz-Institut f\"{u}r Strahlen- und Kernphysik der
Universit\"{a}t Bonn, 53115 Bonn, Germany}
\author{R.~Jahn}
\affiliation{Helmholtz-Institut f\"{u}r Strahlen- und Kernphysik der
Universit\"{a}t Bonn, 53115 Bonn, Germany}

\author{R.~Joosten}
\affiliation{Helmholtz-Institut f\"{u}r Strahlen- und Kernphysik der
Universit\"{a}t Bonn, 53115 Bonn, Germany}

\author{K.~Kilian}
\affiliation{Institut f\"{u}r Kernphysik, Forschungszentrum J\"{u}lich,
52425 J\"{u}lich, Germany}

\author{A.~Kozela}
\affiliation{Institute of Nuclear Physics, Polish Academy of
Sciences, Krakow, Poland}

\author{H.~Machner}
\affiliation{Institut f\"{u}r Kernphysik, Forschungszentrum J\"{u}lich,
52425 J\"{u}lich, Germany}

\author{A.~Magiera}
\affiliation{Institute of Physics, Jagellonian University, 30059
Krakow, Poland}

\author{J.~Munkel}
\affiliation{Helmholtz-Institut f\"{u}r Strahlen- und Kernphysik der
Universit\"{a}t Bonn, 53115 Bonn, Germany}

\author{P.~von~Neumann-–Cosel}
\affiliation{Institut fu¨r Kernphysik, Technische Universit\"{a}t
Darmstadt, 64289 Darmstadt, Germany}

\author{P.~von Rossen}
\affiliation{Institut f\"{u}r Kernphysik, Forschungszentrum J\"{u}lich,
52425 J\"{u}lich, Germany}

\author{H.~Schnitker}
\affiliation{Helmholtz-Institut f\"{u}r Strahlen- und Kernphysik der
Universit\"{a}t Bonn, 53115 Bonn, Germany}

\author{K.~Scho}
\affiliation{Helmholtz-Institut f\"{u}r Strahlen- und Kernphysik der
Universit\"{a}t Bonn, 53115 Bonn, Germany}

\author{J.~Smyrski}
\affiliation{Institute of Physics, Jagellonian University, 30059
Krakow, Poland}
\author{R.~T\"{o}lle}
\affiliation{Institut f\"{u}r Kernphysik, Forschungszentrum J\"{u}lich,
52425 J\"{u}lich, Germany}

\author{C.~Wilkin}
\affiliation{Department of Physics \& Astronomy, UCL, London WC1E
6BT, U.K.}

\collaboration{The COSY-MOMO Collaboration}


\date{\today}

\begin{abstract}%
Two--kaon production in proton--deuteron collisions has been
studied at three energies close to threshold using a calibrated
magnetic spectrograph to measure the final $^3$He and a vertex
detector to measure the $K^+K^-$ pair. Differential and total
cross sections are presented for the production of $\phi$--mesons,
decaying through $\phi\to K^+K^-$, as well as for prompt $K^+K^-$
production. The prompt production seems to follow phase space in
both its differential distributions and in its energy dependence.
The amplitude for the $pd\to\,^3${He}$\,\phi$ reaction varies
little for excess energies below 22\,MeV and the value is
consistent with that obtained from a threshold measurement. The
angular distribution of the $K^+K^-$ decay pair shows that near
threshold the $\phi$--mesons are dominantly produced with
polarization $m=0$ along the initial proton direction. No
conclusive evidence for $f_0(980)$ production is found in the
data.
\end{abstract}


%

\maketitle
\section{Introduction}
The study of meson production in proton--nucleus collisions near
threshold is of interest because of the intricate reaction mechanism
that allows the momentum transfer to be shared among several
nucleons and this feature becomes yet more critical as the mass of
the meson is increased. The simplest reaction of this type is
$pd\to\,^3\textrm{He}\,X^0$, which has the great experimental
advantage that the $^3$He can be detected in a spectrometer and the
meson $X^0$ identified from the missing mass in the reaction. The
cross sections for the near--threshold production of
$\pi^0$~\cite{Nikulin}, $\eta$~\cite{eta},
$\omega$~\cite{Wurzinger95}, and $\eta'$ and
$\phi$~\cite{Wurzinger96} have been extracted in this way. One
drawback of this approach is, however, that in certain cases the
backgrounds from multipion production can be quite large and rapidly
varying. A more intrinsic problem in the case of the production of
the spin--one $\omega$ and $\phi$ mesons is that an inclusive
measurement will contain no information on their polarization. Both
these difficulties can be overcome if products of the decay of the
meson are detected in coincidence with the recoiling $^3$He. The
obvious solution in the $\phi$ case reported here is to measure the
$pd\to\,^3\textrm{He}\,(\phi\to K^+K^-)$ channel which, according to
the Particle Data Group (PDG), has a 49.1\% branching
ratio~\cite{PDG06}. The experiment represents an extension of our
previous work, where we studied two--pion production in the
$pd\to\,^3\textrm{He}\,\pi^+\pi^-$ reaction for excess energies up
to 70\,MeV~\cite{Bellemann99}.

The basic apparatus and how it is used to identify the
$pd\to\,^3\textrm{He}\,K^+K^-$ candidates are described in
Sec.~\ref{Experiment}, with the method of analyzing these events
obtained at three different excess energies being reported in
Sec.~\ref{Data_Analysis}. The separation of $^3\textrm{He}\,\phi$
events from those of prompt $^3\textrm{He}\,K^+K^-$ production is
based principally on the distribution in the $K^+K^-$ invariant
mass. However, it is important to demonstrate that this division
is consistent with the distributions in the other kinematical
variables and this is achieved in Sec.~\ref{Results}. The angular
distributions show evidence for pure $S$--wave production of both
prompt $K^+K^-$ and $\phi\to K^+K^-$ pairs, with the latter being
completely dominated by those where the $\phi$ has polarization
$m=0$ along the beam direction. Only upper limits could be placed
upon the production in this reaction of the $f_0(980)$ scalar
meson decaying into $K^+K^-$. The total cross sections for both
kaon production reactions are given in Sec.~\ref{Discussion},
where it is seen that the energy dependence of prompt and $\phi$
production seem to follow respectively three--body and two--body
phase space. Furthermore, the amplitude for the
$pd\to\,^3\textrm{He}\,\phi$ reaction is consistent with that
obtained at SATURNE using the missing--mass
method~\cite{Wurzinger96}. Our conclusions are summarized in
Sec.~\ref{Conclusions}.
%
%
\section{Experiment}
\label{Experiment}%
\begin{figure}[h]
\begin{center}
\includegraphics[width=8cm]{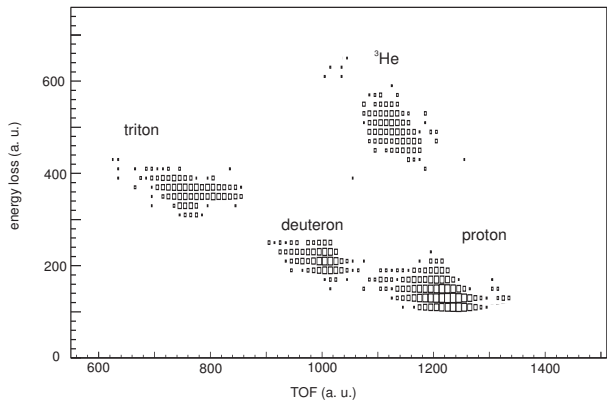}
\caption{Energy loss in the first scintillation layer in the focal
plane of the spectrograph as a function of the time of flight
between the two scintillation hodoscopes, both in arbitrary units.
The well--separated groups of different particles show up very
clearly.} \label{fig:DE_TOF}
\end{center}
\end{figure}

The experiment was carried out at the MOMO (Monitor of Mesonic
Observables) facility, which is installed at an external beam
position of the COSY accelerator of the Forschungszentrum J\"{u}lich.
Near threshold, the $^3$He produced in the
$pd\to\,^3\textrm{He}\,K^+K^-$ reaction go into a narrow forward
cone, where they can be analyzed with the high resolution 3Q2DQ
spectrograph Big Karl~\cite{Drochner98}. Two sets of multiwire drift
chambers (MWDC), placed in the focal plane, were used to measure the
tracks of charged particles. Of these particles, the $^3$He could be
identified unambiguously using two scintillation hodoscopes, placed
downstream of the MWDC and separated by 4\,m, which provided
energy--loss and time--of--flight information. The effectiveness of
this approach is illustrated in Fig.~\ref{fig:DE_TOF}, where it is
seen that different particle types show up as well--separated
groups.

\begin{figure}
\begin{center}
\includegraphics[width=8 cm]{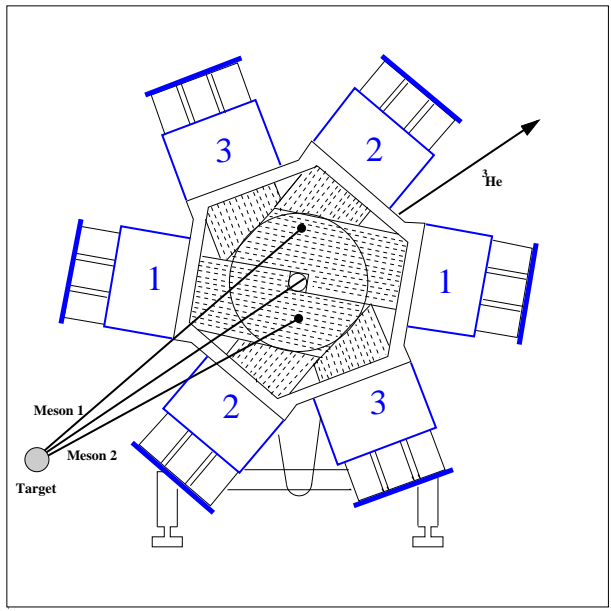}
\caption{Front view to the MOMO vertex detector with the indication
of a typical event. Both the primary beam and the recoil $^3$He
detected in Big Karl pass through the central hole. The numbers
denote the different layers and the three boxes at the end of each
read--out symbolize the phototubes. The support stand is visible in
the lower part of the figure.} \label{Fig:MOMO_vertex}
\end{center}
\end{figure}

Two charged kaons were measured in coincidence with the ${^3}$He
ions using the MOMO vertex detector. This combination had been
proved to work successfully for two--pion production in
Ref.~\cite{Bellemann99}. The vertex detector, a schematic view of
which is shown in Fig.~\ref{Fig:MOMO_vertex}, consists of three
layers of 2.5\,mm thick scintillating fibers with 224 parallel
fibers in each layer. These are rotated by 60$^{\circ}$ to each
other with read--outs through phototubes on opposite sides. The
detector is placed outside the vacuum chamber containing the target,
some 20\,cm downstream of the target. Hits in three layers are
required in order to avoid the combinatorial ambiguities associated
with two hits.

In contrast to the near--threshold two--pion production
experiment~\cite{Bellemann99}, the multipion background is very
large for effective masses in the GeV/c$^2$ region. In view of this,
and in order to identify the produced particles unambiguously as
kaons, the detector was supplemented by a hodoscope consisting of 16
wedge--shaped scintillators, each 2\,cm thick, the details being
given in Ref.~\cite{theses}. This modified vertex detector was
calibrated with events from elastic $pp$ scattering. Charged kaons
could thus be detected and their production vertex measured with
full azimuthal acceptance within a polar angular range of
$8^{\circ}<\theta_{\rm lab}<45^{\circ}$. This modified vertex
detector was calibrated with events from elastic $pp$ scattering. It
is important to note that, since there is no magnetic field
associated with the MOMO detector, it is not possible to identify
the charge of an individual kaon and this automatically symmetrizes
some of the distributions.

The liquid deuterium target was a cylinder of diameter 6\,mm and
4\,mm thickness with $0.9\,\mu$m mylar windows~\cite{Jaeckle94}. The
small beam diameter of less than 2\,mm led to a precise
determination of the emission directions, i.e. polar and azimuthal
angles. The incident beam intensity was monitored by calibrated
scintillators which, on the basis of the known target areal density,
allowed the absolute cross sections to be
evaluated~\cite{Drochner98}.

\begin{figure}[h]
\centering
\includegraphics[width=8 cm]{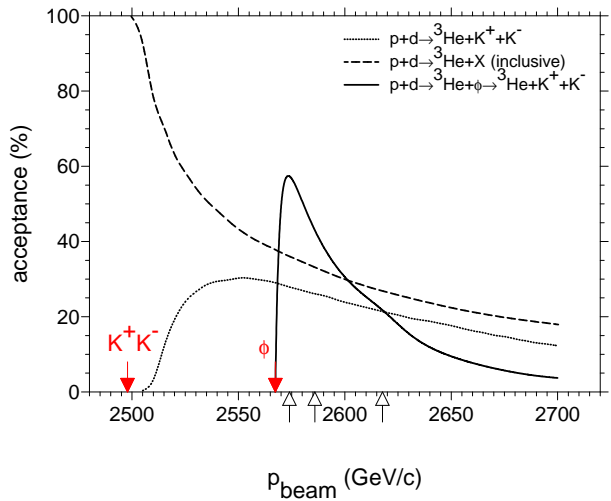}
\caption{The efficiency of the apparatus to accept inclusive
$^3$He production (dashed curve), prompt $K^+K^-$ production
integrated over the two--kaon excitation energy (chain curve), and
$\phi$ production through the $K^+K^-$ channel, as functions of
the proton beam momentum. The thresholds for $K^+K^-$ and $\phi$
production are indicated while the momenta employed in this work
are shown by arrows with open heads.} \label{fig:acceptance}
\end{figure}

Monte Carlo estimates of the overall efficiency of the apparatus to
detect the $^3$He$\,K^+K^-$ and $^3$He$\,\phi$ final states are
shown in Fig.~\ref{fig:acceptance} as functions of the beam
momentum. Big Karl has a momentum bite of $\pm 4\%$ of the central
momentum. For this momentum it has a horizontal and vertical
acceptance of $\pm25\,$mrad and $\pm100\,$mrad about the beam
direction and smaller acceptance for the other momenta.  Therefore a
little above threshold all the $^3$He from inclusive
$^3$He$\,K^+K^-$ production should be covered. However, the total
system is blind under such conditions since the emitted kaons fall
within the central hole of the MOMO detector and are lost. On the
other hand, near its threshold, the acceptance for the $\phi$ is
high, because the transverse momentum of the $^3$He is low while the
kaon opening angle is comparatively large.

The experiment was carried out at three overall excess energies
$\varepsilon_{KK}=35.1$, 40.6, and 55.2\,MeV, \emph{i.e}\ excess
energies in the $^3$He$\,\phi$ system of $\varepsilon_{\phi}=3.0$,
8.5, and 23.1\,MeV. The corresponding beam momenta are marked on
Fig.~\ref{fig:acceptance}, from where it can be seen that the
efficiency for $\phi$ detection was always greater than 20$\%$.
Also marked there are the $K^+K^-$ and $\phi$ thresholds, though
the latter is made fuzzy by the meson's $(4.26\pm0.05)$\,MeV/c$^2$
width~\cite{PDG06}. It should be stressed that kaons with momenta
below 160\,MeV/c are stopped within the end wall of the MOMO
scattering chamber so that the probability of two kaons reaching
the detector is always larger than $70\%$.

The limited momentum bite of the spectrograph meant that between
three and five settings of its central value were required for each
beam momentum in order to cover the full phase space of the
reaction. The integrated luminosity was typically $2\times
10^{36}/$\,cm$^{-2}$ per setting. Runs were performed with an empty
target cell in order to study the background. The events recorded
under these conditions were analyzed in the same way as those from
the target--full runs. The fraction of background events was only
0.2\% and these were subtracted from the data sample.

%
%
\section{Data Analysis}
\label{Data_Analysis}%
\begin{figure}[h]
\centering
\includegraphics[width=8 cm]{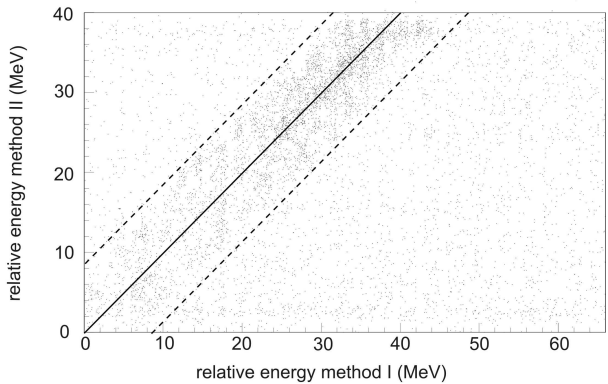}
\caption{The excitation energy in the $KK$ system obtained in the
$\varepsilon_{KK} = 40.6\,$MeV run, evaluated I from the invariant
$KK$ mass, and II from the missing mass of the $^3$He measured
with Big Karl.} \label{fig:TK1_TK2}
\end{figure}

Having identified $^3$He$\,K^+K^-$ candidates on the basis of the
spectrograph and MOMO information, much of the background could be
eliminated by demanding that the events be coplanar in the cm
system. The measurement of the $^3$He momentum together with the
kaon directions means that one has a two--constraint fit to the
reaction and this reduces the uncertainties in both the
identification of the reaction and of its kinematics. As an example
of this, we show in Fig.~\ref{fig:TK1_TK2} the $KK$ excitation
energy, obtained from the missing--mass in Big Karl, plotted against
that deduced from the invariant mass extracted using the
reconstructed kaon momenta. Good events lie along the diagonal and
we retain those within $\pm 9\,$MeV/c$^2$ of the value expected from
the spectrograph measurement.

The corrections necessary for the extraction of the differential
cross sections depend upon the particular angle and energy bin as
well as on the spectrograph setting. The efficiency for resonant
two--kaon production \emph{via} the $\phi$ is therefore different
from that of the prompt production integrated over the $K^+K^-$
excitation energy.

The luminosity measurement needed to derive absolute cross sections
has systematic uncertainties of 5$\%$ from the target thickness and
5$\%$ from the beam intensity. Another systematic uncertainty stems
from the efficiency correction, which ranges from 5\% up to 20\% for
small values of the excitation energy in the $K^+K^-$ system. With a
beam intensity of $\leq 10^{9}$s$^{-1}$, the dead time was
negligible. The numbers of identified events were transformed to
cross sections, taking the detector efficiencies into account.

%
%
\section{Differential Distributions}
\label{Results}%
\begin{figure}[h]
\begin{center}
\includegraphics[width=7 cm]{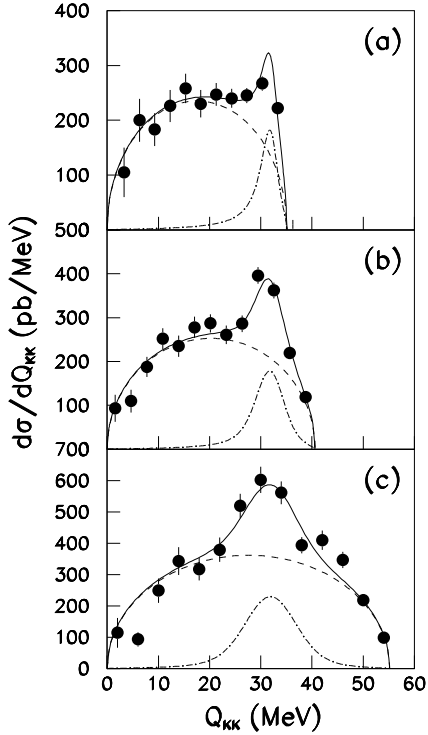}\vspace{-0.5 cm}
\caption{\label{fig:T_KK}Cross section for the reaction $pd\to\,
^3\textrm{He}\,K^+K^-$ as a function of the excitation energy
$Q_{KK}$ of the two kaons at overall excess energies of (a)
$\varepsilon_{KK}=35.1\,$MeV, (b) 40.6\,MeV, and 55.2\,MeV. The
events are binned in equally spaced intervals. The curves are fits
to the $Q_{KK}$ distributions in terms of phase space coming from
prompt $K^+K^-$ production (dashed line), proceeding \emph{via}
$\phi$--meson formation (chain), and their sum (solid line). }
\end{center}
\end{figure}

\begin{figure}[h]
\begin{center}
\includegraphics[width=8 cm]{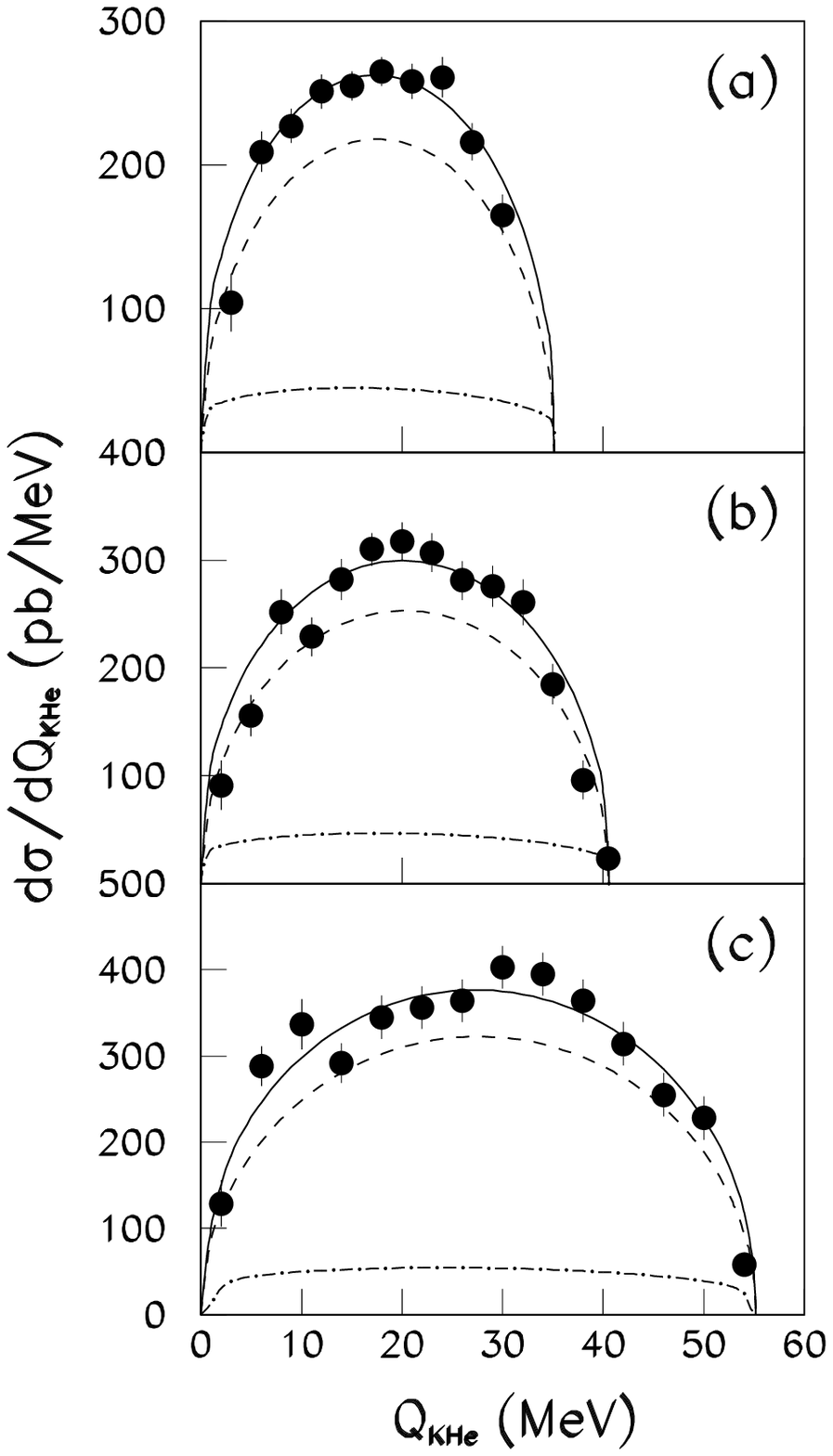}\vspace{-1cm}
\caption{\label{fig:T_KHe}Same as Fig.~\ref{fig:T_KK} but as a
function of the excitation energy $Q_{KHe}$ in the $K\,^3$He
system. The consequences of the fits shown in Fig.~\ref{fig:T_KK}
for the $Q_{KHe}$ distributions are shown, though the deviations
from simple three--body phase space are here relatively minor.
This is due in part to the averaging over the $K^-\,^3$He and
$K^+\,^3$He distributions by the MOMO apparatus.}
\end{center}
\end{figure}

In Figs.~\ref{fig:T_KK} and \ref{fig:T_KHe} we show the projections
of the Dalitz-like plot for the $pd\to\,^3\textrm{He}\,K^+K^-$
reaction at the three excess energies onto axes corresponding to the
excitation energies $Q_{KK}$ and $Q_{KHe}$ in the $K^+K^-$ and
$K\,^3$He systems respectively. Since the value of $Q_{KK}$ is fixed
completely by the measurement with the high resolution spectrograph
Big Karl, the distribution in this variable is the best determined
of all the differential cross sections. All the other distributions
rely primarily on the information furnished by the vertex detector,
though the spectrograph data refines these through the kinematic
fitting.


There is evidence for the production of the $\phi$ meson at all
three energies but the physics background arising from a prompt
$K^+K^-$ production looking like phase space is very large. This
behavior of the prompt $K^+K^-$ pairs is very different from that
found for the $pd\to\,^3\textrm{He}\,\pi^+\pi^-$ reaction, studied
with the same apparatus~\cite{Bellemann99}, where effects from
double $p$--wave production are very evident.

Fits to the excitation energy spectra have been undertaken in terms
of phase space and phase space modulated by a $\phi$ peak, which has
been taken to have a Breit--Wigner form with a natural width
$\Gamma=4.2\,$MeV/c$^2$~\cite{PDG06}. This has been folded with an
energy resolution width $\sigma$, which reflects uncertainties in
the overall system, including the beam momentum spread, as well as
effects arising from the binning of the data. The predictions of
these fits are shown in Fig.~\ref{fig:T_KK}, with their reflections
on the $K\,^3$He spectra being presented in Fig.~\ref{fig:T_KHe}.
The broad agreement achieved here supports the basic \emph{ansatz}
that the only distortion of phase space is that due to the $\phi$
peak. In addition to kinematic effects, the broader $\phi$ peak at
55.2 MeV is due in part to the less favorable beam conditions.

\begin{figure}[h]
\begin{center}
\includegraphics[width=8 cm]{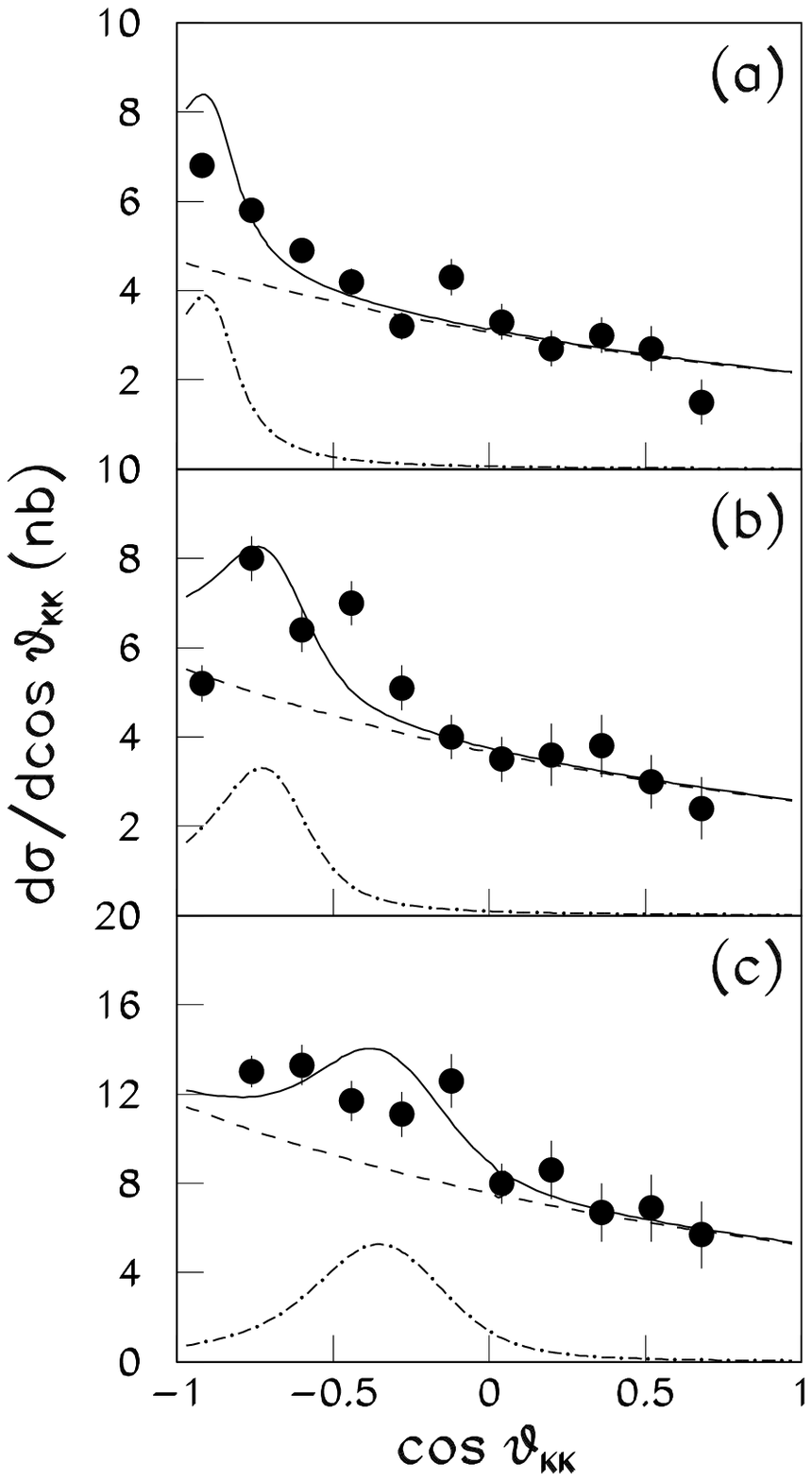}\vspace{-1cm}
\caption{Distribution in the cosine of the opening angle
$\theta_{KK}$ at (a) $\varepsilon_{KK}=35.1$\,MeV, (b) 40.6\,MeV,
and 55.2\,MeV. The dashed line represents the phase-space
contribution arising from prompt $K^+K^-$ production, whereas the
chain corresponds to the $\phi$ component and the solid line the sum
thereof.} \label{fig:kaon_angles}
\end{center}
\end{figure}
\begin{figure}[h]
\begin{center}
\includegraphics[width=8 cm]{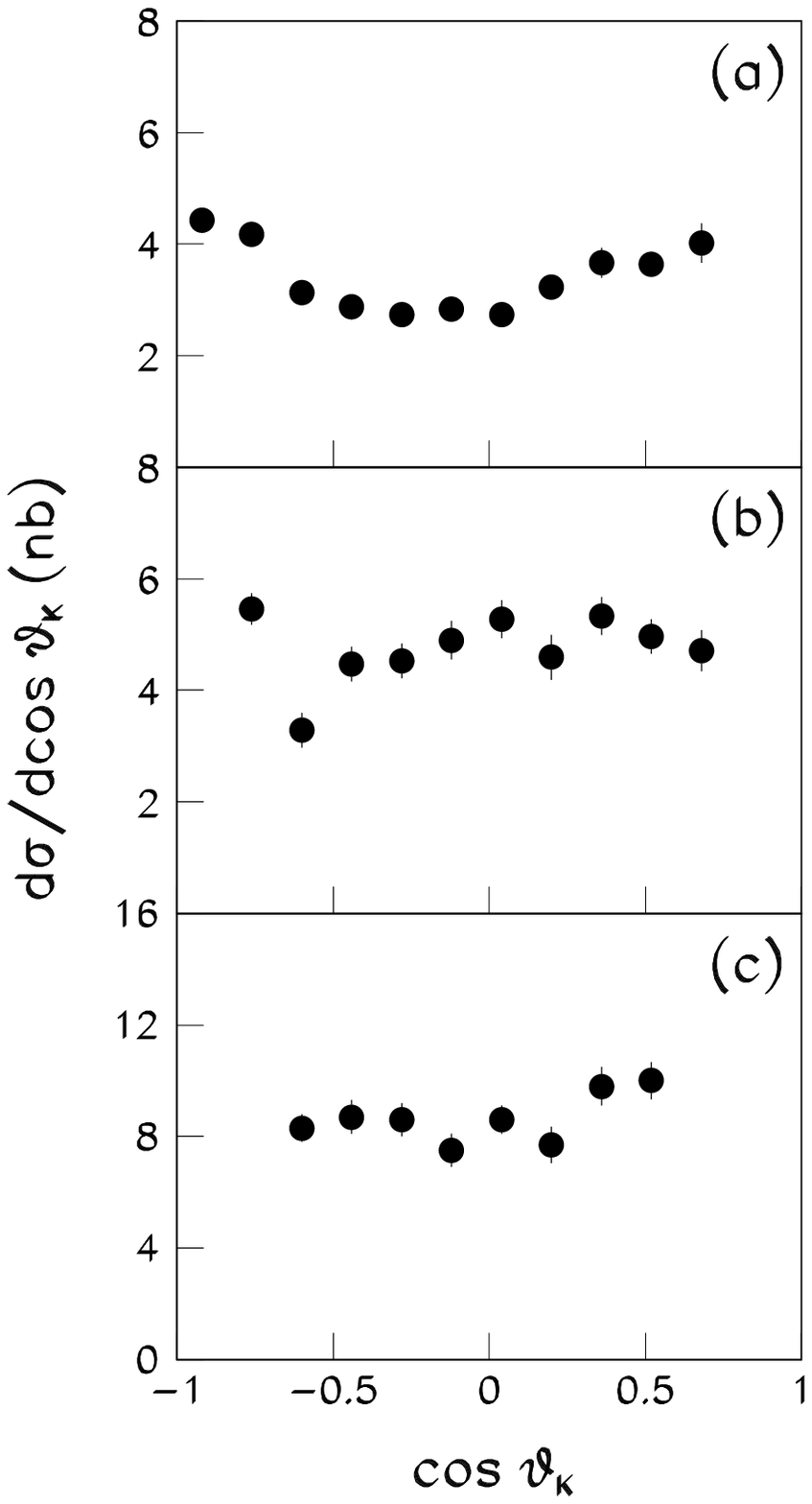}\vspace{-1cm}
\caption{Same as Fig.~\ref{fig:kaon_angles} but for the
distribution in the angle between an outgoing kaon and the
incident proton in the overall cm system. For prompt production of
$S$--wave kaon pairs, this distribution is expected to be
isotropic.  However, a variation with angle can be generated
through a polarization dependence of $\phi$ production.}
\label{fig:kaon_he}
\end{center}
\end{figure}

Within the present statistics, the differential cross sections for
both the prompt $K^+K^-$ and $\phi$ emission are independent of
the $^3$He cm angle, as expected for $S$--wave production.
Fig.~\ref{fig:kaon_angles} shows the variation of the cross
section as a function of the opening angle between the two kaons.
The phase--space distribution is expected to be about $2.2$ times
bigger in the backward direction than in the forward. This is a
purely kinematic effect, as is the peak arising from $\phi$
production. The decay kaons are emitted back to back in the $\phi$
rest system, which coincides with the overall cm system at the
$\phi$ threshold. At the lowest energy there is therefore an
enhancement close to $\cos\theta_{KK}=-1$ but, for higher excess
energy, this is shifted towards smaller opening angles due to the
random orientation of the $\phi$ decay products with respect to
the $\phi$ momentum.

Figure~\ref{fig:kaon_he}  shows the distribution in the angle
between one of the kaons and the beam axis in the overall cm
system. Pure phase--space would lead to isotropy but even the
$S$--wave production of the $\phi$ meson can lead to some
dependence on $\theta_K$ if the $\phi$--meson is produced
polarized. This can be seen more transparently in the
Gottfried--Jackson frame~\cite{GJ}, described below. This
polarization effect also decreases as the beam energy is raised,
because of the random orientation of the produced $\phi$ meson in
the cm frame.

\begin{figure}[h]
\begin{center}
\includegraphics[width=8cm]{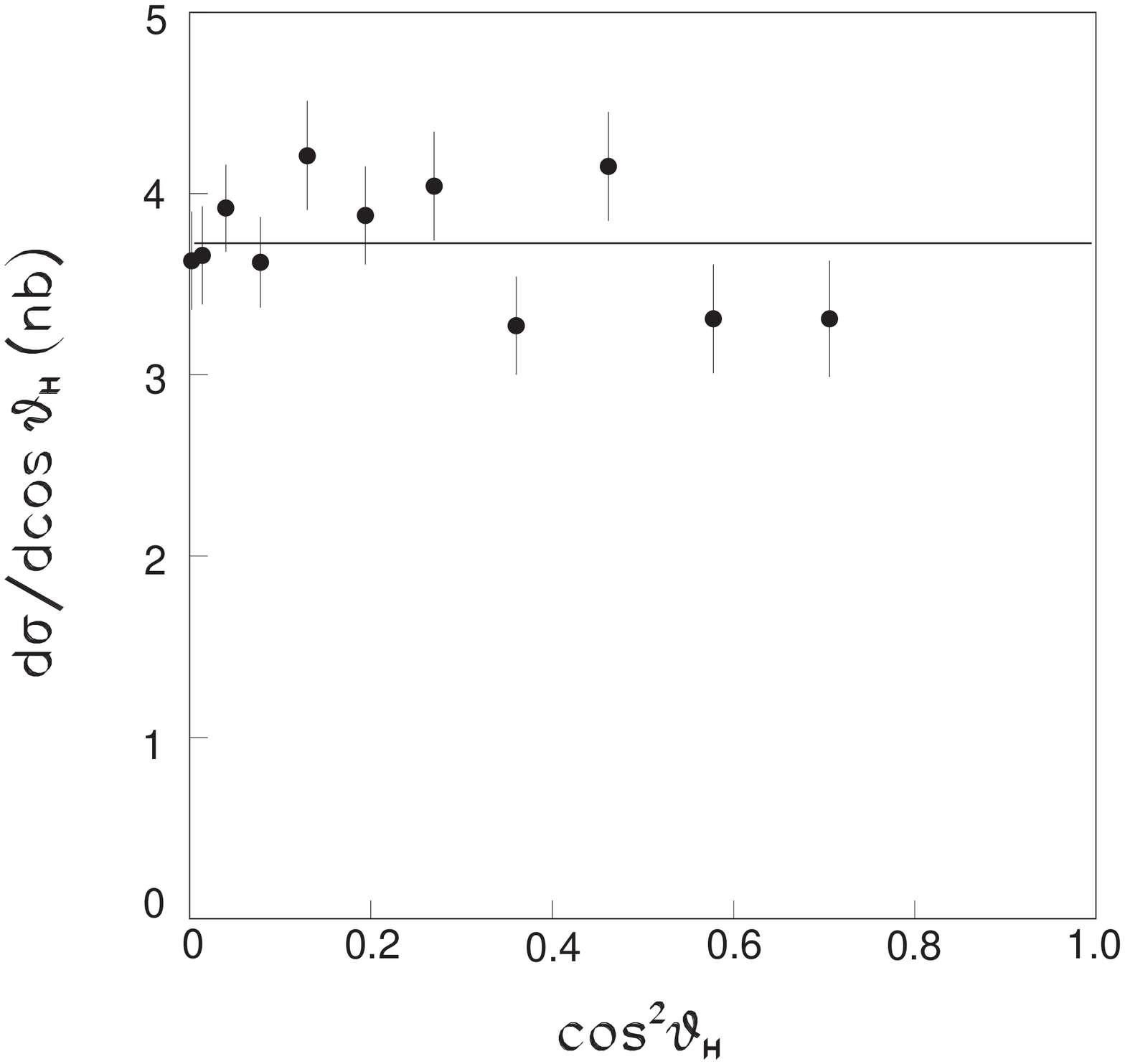}\vspace{-0.0 cm}
\caption{Angular distributions of the kaons as a function of the
helicity angle for the lowest energy of
$\varepsilon_{KK}=35.1\,$MeV. The data are consistent with the
average value indicated.}
 \label{fig:helicity_frame}
\end{center}
\end{figure}

In the Gottfried-Jackson frame~\cite{GJ,Byck73}, the total
momentum of the $K^+K^-$ system is zero, which means that it is
the $\phi$ rest frame. Since the $\phi$ is a vector meson, the
distribution in the relative momentum of the kaons from its decay
is sensitive to its polarization with respect to some quantization
axis:
\begin{equation}\label{equ:density_matrix}
\frac{\rd\sigma(\phi\to 2K)}{\rd\cos (\theta_{GJ})}\propto
\rho_{11}\sin^2\theta_{GJ}+\rho_{00}\cos^2\theta_{GJ}.
\end{equation}
Here the spin-density matrix elements $\rho_{00}$ and $\rho_{11}$
correspond to the populations with magnetic sub--state $m=0$ and
the average of $m=\pm 1$ respectively. On the other hand, the
production of an $S$--wave $K^+K^-$ pair would lead to a flat
distribution in the decay angle $\theta_{GJ}$.

The helicity distribution is obtained by choosing the quantization
axis to lie along that of the recoiling $^3$He nucleus. Since any
anisotropy here must be proportional to the square of the $^3$He
momentum, \emph{i.e.}\ the excess energy in the $\phi\,^3$He
system, it is not surprising that the results shown in
Fig.~\ref{fig:helicity_frame} at $\varepsilon_{\phi}=3\,$MeV are
consistent with a flat distribution.

\begin{figure}[h]
\begin{center}
\includegraphics[width=8cm]{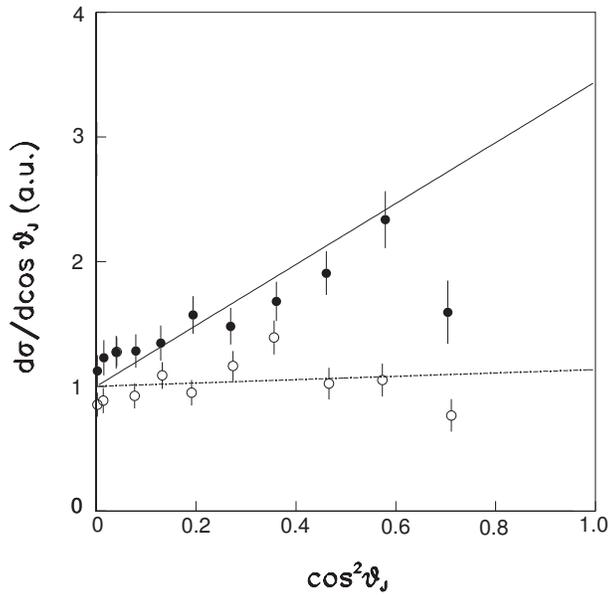}\vspace{0.5 cm}
\caption{Angular distributions of the kaons as a function of the
Jackson angle for the lowest energy of
$\varepsilon_{KK}=35.1\,$MeV. The data are divided into those with
$Q_{KK}<28\,$MeV (closed circles) and $Q_{KK}>28\,$MeV (open
circles). In order to show clearly the difference in slopes, the
data have been arbitrarily scaled so that they have similar values
at the origin. The straight lines shown here are predictions based
on the fits to Fig.~\ref{fig:T_KK}, assuming that the $\phi$ are
produced only with $m=0$ along the proton beam direction.}
\label{fig:GJ_frame}
\end{center}
\end{figure}

The axis for the Jackson angle is taken to be the relative
momentum in the initial system which, for near--threshold
production, can be replaced by the incident proton momentum. In
the right panel of Fig.~\ref{fig:GJ_frame} the distribution in
this angle is shown separately for the $\phi$--rich region, where
$Q_{KK}>28\,$MeV, and the remainder. In order to demonstrate the
very different slopes in the two regions, the data have been
arbitrarily scaled such that the cross sections have similar
values when $\cos\theta_{GJ}=0$. These slopes are determined by
the fraction of the cross section associated with $\phi$
production and the $\phi$ polarization. The straight lines in the
figure are obtained by using the $\phi$ contribution determined
from the fits to the $Q_{KK}$ distribution of Fig.~\ref{fig:T_KK}
assuming that the meson is produced purely with $m=0$. This
\emph{ansatz} describes the main features of the data in both
energy regions. Alternatively, fitting
Eq.~(\ref{equ:density_matrix}) to the data with $Q_{KK}>28\,$MeV
gives $\rho_{00}=0.82\pm0.05$, where the error bar is statistical
and does not take into account that arising from the
identification of the $\phi$ cross section.

The data of Fig.~\ref{fig:T_KK} may be used to try to put limits on
the cross section for scalar meson production and, in particular, on
the $f_0(980)$. Whereas PDG reports that the decay
$f_0(980)\to{}K\bar{K}$ is merely \emph{seen}, a recent measurement
of the decay into pions and kaons performed at BES2~\cite{BES}
yielded a $\displaystyle (25^{+11}_{-13})\%$ branch into $K\bar{K}$.
We re-analyzed the $Q_{KK}$ distribution adding incoherently a third
component corresponding to a state of mass 980\,MeV/c$^2$ and width
$\Gamma=47$\,MeV/c$^2$. The fits yielded zero cross section for the
$K^+K^-$ branch with upper limits of 6\%, 7\%, and 9\% of the prompt
$K^+K^-$ cross section at our three energies.
\begin{figure}[h]
\begin{center}
\includegraphics[width=8 cm]{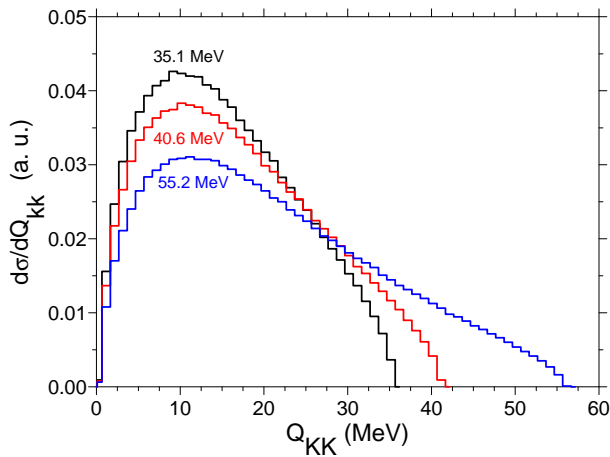}
\caption{Shape of the cross section expected for the reaction
$pd\to\,^3$He$\,f_0(980)$, with the subsequent decay $f_0(980)\to
K^+K^-$, for the excess energies indicated. All predictions have
been normalized to unity.} \label{fig:f0}
\end{center}
\end{figure}

In Fig.~\ref{fig:f0} we show the shape expected for a $f_0(980)$
contribution to the differential spectrum at the three excess
energies. The maximum around $Q_{KK}\approx 10$\,MeV spoils the
agreement with the shape of the experimental spectra. However,
typically only three data points are in this range and it is
precisely here that the uncertainties in the efficiency
corrections are the largest. Higher statistics data would be
needed to pin down unambiguously the fraction of scalar meson
production in the present reaction. Such data might be obtained
from studies of the stronger decay channel into two pions but the
angular acceptance of the MOMO vertex detector is too small to
detect these particles.
%
%
\section{Total Cross Sections}
\label{Discussion}%

The total $\phi$ and prompt kaon production cross sections, obtained
by integrating the fits to Fig.~\ref{fig:T_KK}, are presented in
Table~\ref{results}. In the $\phi$ case, the branching ratio
BR$(\phi\to K^+K^-)=49.1\%$~\cite{PDG06} has been included. The
value of the energy resolution parameter $\sigma$ is that deduced
from the $\phi$ peak, after taking the natural width of
$4.2\,$MeV/c$^2$ into account.

\begin{table*}[hbt]
\caption{Total cross sections for prompt $K^+K^-$ and $\phi$
production in terms of the incident beam momentum and their
respective excess energies $\varepsilon_{KK}$ and
$\varepsilon_{\phi}$; the $\phi$ results have been corrected for the
49.1\% charged--kaon branching ratio.  The square of the average
$pd\to\,^3\textrm{He}\,\phi$ amplitude is extracted from the total
$\phi$ production cross section through
Eq.~(\ref{equ:Matrix-element}). In addition to the statistical
errors, the quoted include also those arising from measurements of
the beam intensity and uncertainties in the acceptance correction
and the separation of the $\phi$ peak from the non resonant kaon
production. Not shown is the overall uncertainty of $\pm5\%$ in the
areal density of the target. } \label{results}
\begin{ruledtabular}
\begin{tabular}{lccc}
Beam Momentum (MeV/c) & $2574\pm1$ & $2586\pm1$ &
$2618\pm2$\\
\hline
$\varepsilon_{KK}$ (MeV)& $35.1\pm 0.5$ & $40.6\pm0.5$&$55.2\pm0.8$\\
$\sigma_{KK}$ (nb)& $6.4\pm 0.5$ & $8.1\pm0.5$ & $15.8\pm 1.0$\\
\hline
$\varepsilon_{\phi}$ (MeV) & $3.0\pm0.5$&$8.5\pm0.5$&$23.1\pm0.8$\\
$\sigma_{\phi}$ (nb)&$2.0\pm0.4$&$3.0\pm0.6$&$6.4\pm1.8$\\
$|f|^2$ (nb/sr) &$3.0\pm0.6$&$3.1\pm0.6$&$3.4\pm1.0$\\
\end{tabular}
\end{ruledtabular} \vspace{0.5 cm}
\end{table*}

\begin{figure}[h]
\centering
\includegraphics[width=8 cm]{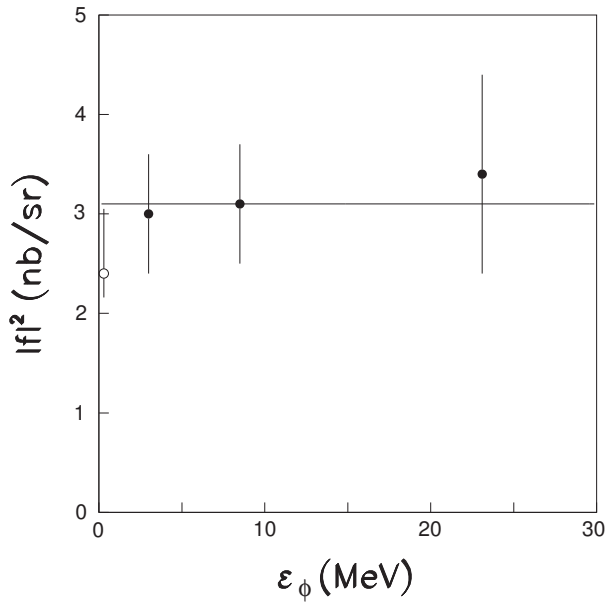}
\caption{The variation of the amplitude--square for $\phi$
production, as defined by Eq.~(\ref{equ:Matrix-element}), with
$\varepsilon_{\phi}$. The open circle at
$\varepsilon_{\phi}=0.3\,$MeV is taken from the SPESIV
measurement~\cite{Wurzinger96}. The results are all consistent with
$|f|^2$ being constant, as indicated.} \label{fig:matrix}
\end{figure}

The spin--averaged square of the matrix element for $\phi$
production can be extracted from the total cross section using
\begin{equation}\label{equ:Matrix-element}
|f|^2=\frac{1}{4\pi}\,\frac{p_p}{p_{He}}\,
\sigma_T(pd\to\,^3\textrm{He}\,\phi)\:,
\end{equation}
where $p_p/p_{He}$ is the phase--space factor of the ratio of the
incident to the final cm momenta. Provided that there is no strong
interaction between the $\phi$ and the $^3$He, one would expect at
these low energies to have predominantly $S$--wave production with
very little variation of $|f|^2$, and this is what is seen from the
results shown in Table~\ref{results}. Furthermore, the values
obtained are consistent with that found with the SPESIV
spectrometer~\cite{Wurzinger96}, which at
$\varepsilon_{\phi}=0.3\,$MeV gave
$|f|^2=(2.4\pm0.2\,^{+0.6}_{-0.2})\,$nb/sr, where the first error is
statistical and the second systematic. The mean value of this and
our results, which are shown together in Fig.~\ref{fig:matrix}, is
$(3.0\pm0.2)\,$nb/sr. This very smooth behavior with energy is to be
contrasted with the case of $pd\to\,^3\textrm{He}\,\omega$, where
the amplitude is seen to be suppressed as soon as the excess energy
is similar to the width of the $\omega$~\cite{Wurzinger95}, though
the interpretation of these data has been questioned~\cite{HK}.

\begin{figure}[h]
\centering
\includegraphics[width=8 cm]{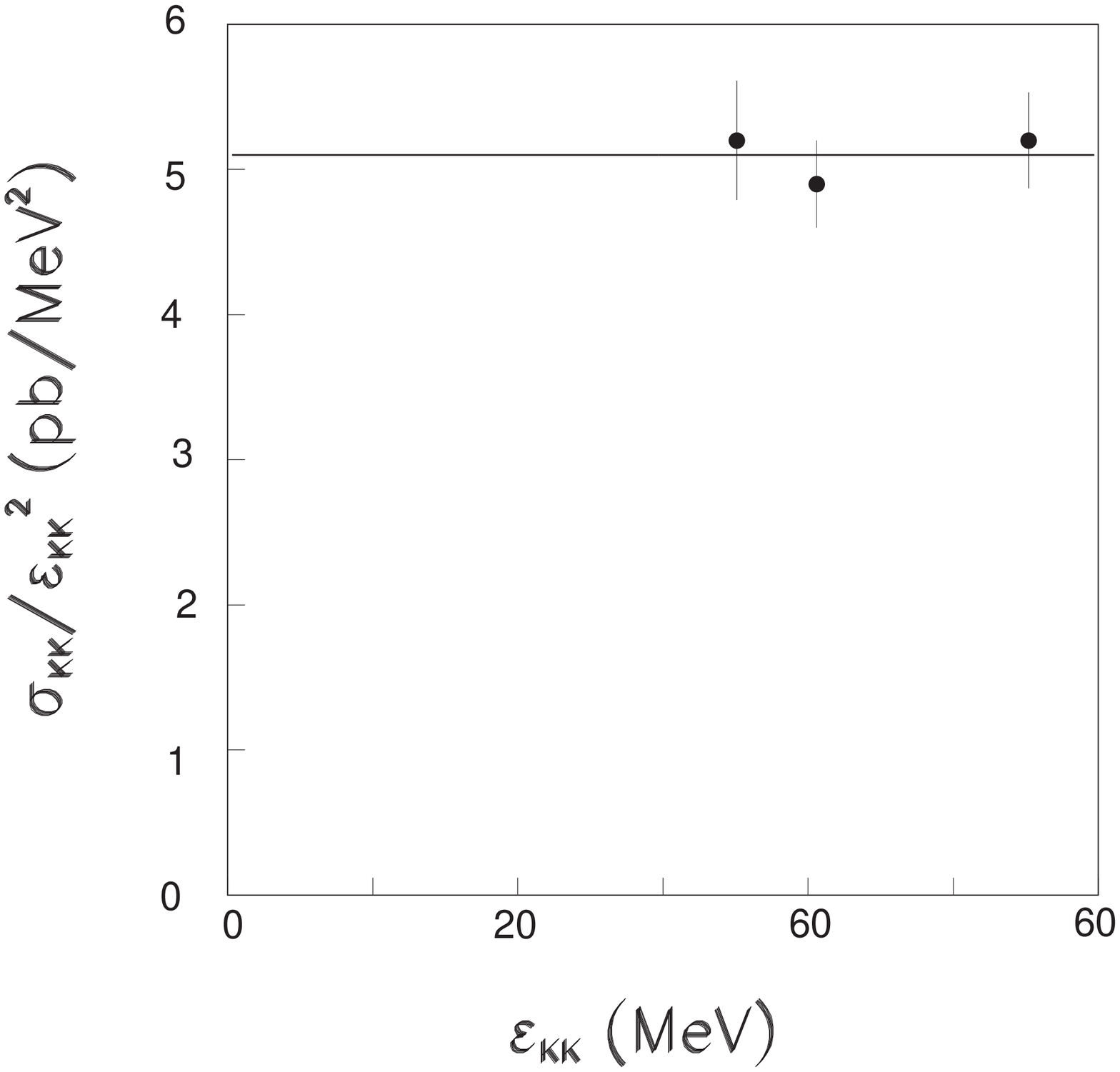}\vspace{0.0 cm}
\caption{Total cross section for prompt two--kaon production divided
by a phase--space factor of $\varepsilon_{KK}^2$ as a function of
$\varepsilon_{KK}$. The average value is also indicated.}
\label{fig:exfu}
\end{figure}

If the prompt $K^+K^-$ production is not influenced by resonances
or other dynamical effects, then the total cross section
$\sigma_{KK}$ might be expected to vary like phase space,
\emph{i.e.}\ as $\varepsilon_{KK}^2$. The ratio
$\sigma_{KK}/\varepsilon_{KK}^2$ shown in Fig.~\ref{fig:exfu} is
consistent with the constant value of $(5.0\pm0.2)\,$pb/MeV$^2$.
The absence of any obvious effects from the $S$--wave $a_0/f_0$
resonances, both in the total $K^+K^-$ production cross sections
and in the $Q_{KK}$ distributions of Fig.~\ref{fig:T_KK}, could be
due to their very large widths and small branching
ratios~\cite{PDG06}.

The only dynamical estimate of $\phi$ production in this reaction
has been made in a two--step model, where a pion beam is produced
\emph{via} $pp\to d\pi^+$ on one target nucleon, with the $\phi$
meson being created through a subsequent $\pi^+n\to p\,\phi$
reaction on the second nucleon in the target. Though this approach
reproduces reasonably well the rates for $\eta$, $\omega$, and
$\eta'$ production, it underpredicts $\phi$ production by at least
a factor of five~\cite{Faeldt95}. Other theoretical models are
therefore necessary to describe $\phi$ production, possibly
involving intermediate $\rho$ as well as $\pi$ mesons.

The ratio of $\phi$ to $\omega$ production in various nuclear
reactions has often been discussed in terms of the OZI
rule~\cite{OZI}, which suggests that, due to $\omega/\phi$ mixing
at the quark level, the ratio
\begin{equation}
\label{OZIrat}
R_{\phi/\omega}\equiv\frac{\sigma_T(pd\to\,^3\textrm{He}\,\phi)}
{\sigma_{T}(pp\to\,^3\textrm{He}\,\omega)}
\end{equation}
should be of the order of $R_\mathrm{OZI}=4.2\,\times 10^{-3}$. As
discussed by Wurzinger et al.~\cite{Wurzinger96}, the difficulty
in extracting numerical values for this ratio resides in the very
strong and unexplained energy dependence observed in the amplitude
for $\omega$ production~\cite{Wurzinger95}. If we follow their
prescription to correct for this and other effects, we find that
$R_{\phi/\omega}\approx 20\times R_\mathrm{OZI}$, which is much
larger than the ratio obtained in near--threshold production in
proton--proton collisions at similar excitation
energies~\cite{Hartmann}.

%
%
\section{Conclusions}
\label{Conclusions}%
In summary, we have made exclusive measurements of the
$pd\to\,^3\textrm{He}\,K^+K^-$ reaction at three energies above
the $\phi$ threshold. By making fits to the $K^+K^-$ excitation
energy distribution in terms of phase space plus a resonance
contribution, we have decomposed the cross section into terms
corresponding to prompt $K^+K^-$ and $\phi$ production.
Distributions in other variables seem to be consistent with this
assumption and no firm evidence is found for the scalar resonance
$f_0(980)$ decaying into $K^+K^-$. Both data sets are consistent
with pure $S$--wave production, with the $K^+K^-$ cross section
varying like $\varepsilon_{KK}^2$ and the $\phi$ as
$\varepsilon_{\phi}^{1/2}$. The most striking effect though comes
from the study of the decay distribution in the $\phi$ rest frame,
which shows that the $\phi$ is formed predominantly with
polarization $m=0$ along the proton beam direction, and this must
be an important clue to the dynamics. Data on the analogous
$pd\to\,^3\textrm{He}\,\omega$ reaction are currently being
analyzed at CELSIUS~\cite{Karin}. The polarization of the $\omega$
is measured through the $\phi\to \pi^0\pi^+\pi^-$ decay and the
results could be particularly illuminating.\vspace{0.65cm}

%
%
\section*{ACKNOWLEDGEMENTS}

\vspace{-1.8mm} The quality of the beam necessary for the success
of this work is due mainly to the efforts of the COSY operator
crew. Thanks goes also to Big Karl technical staff for continuous
help. Support by Forschungszentrum J\"ulich (FFE) and
Bundesministerium f\"{u}r Forschung und Wissenschaft is gratefully
acknowledged.

%
%

\end{document}